\documentclass[aps,prb,groupedaddress,superscriptaddress,preprint]{revtex4}

\usepackage{graphicx}
\usepackage{dcolumn}
\usepackage{bm}

\usepackage{color}

\begin{document}
\title{Simulation of the enhanced Curie temperature in Mn$_5$Ge$_3$C$_x$ compounds}
\author{I. Slipukhina}
\affiliation{Laboratoire de simulation atomistique (L\_Sim), SP2M, INAC, CEA, 38054 Grenoble cedex 9, France}
\author{E. Arras}
\affiliation{Laboratoire de simulation atomistique (L\_Sim), SP2M, INAC, CEA, 38054 Grenoble cedex 9, France}
\author{Ph. Mavropoulos}
\affiliation{Institut f\"ur Festk\"orperforschung and Institute for Advanced Simulation, Forschungszentrum J\"ulich,  D-52425 J\"ulich, Germany}
\author{P. Pochet}
\affiliation{Laboratoire de simulation atomistique (L\_Sim), SP2M, INAC, CEA, 38054 Grenoble cedex 9, France}

\date{February 17, 2009}
\begin{abstract}

Mn$_5$Ge$_3$C$_x$ films with $x\ge0.5$ were experimentally shown
to exhibit a strongly enhanced Curie temperature $T_\mathrm{C}$
compared to Mn$_5$Ge$_3$. In this letter we present the results of our
first principles calculations within Green's function approach,
focusing on the effect of carbon doping on the electronic and magnetic
properties of the Mn$_5$Ge$_3$. The calculated exchange coupling
constants revealed an enhancement of the ferromagnetic Mn-Mn
interactions mediated by carbon. The essentially increased
$T_\mathrm{C}$ in Mn$_5$Ge$_3$C is well reproduced in our Monte Carlo
simulations and together with the decrease of the total magnetisation
is found to be predominantly of an electronic nature.

\end{abstract}

\maketitle
\label{sec:introduction}

Intense efforts have recently been devoted to the fabrication of
high-$T_{\mathrm{C}}$ metal/metalloid materials for their potential
incorporation into spintronic devices. Among them,
Mn$_5$Ge$_3$ seems to be a promising candidate due to its
compatibility with mainstream silicon
technology~\cite{Picozzi2004}. The low Curie temperature
($T_\mathrm{C}$ = 304~K \cite{Tawara1963}) of this compound, which is
a main disadvantage for technological applications, has recently been
overcome by carbon doping in Mn$_5$Ge$_3$C$_x$
films.\cite{Gajdzik2000} It was shown\cite{Gajdzik2000} that carbon
incorporation into the octahedral voids of the hexagonal Mn$_5$Ge$_3$
cell leads to a continous increase of $T_{\mathrm{C}}$ with the doping
level in Mn$_5$Ge$_3$C$_x$ and suggests a saturation when all
octahedral voids are filled by carbon. A maximum $T_\mathrm{C}$ of
445~K was obtained for Mn$_5$Ge$_3$C$_{0.8}$ films, obtained by
magnetron sputtering. The carbon implanted Mn$_5$Ge$_3$C$_{0.8}$ films
were found to exhibit magnetic properties very similar to their
sputtered counterparts~\cite{Surgers2008}. The average saturated
moment of 2.2~$\mu_\mathrm{B}/$Mn was observed for C-implanted
Mn$_5$Ge$_3$C$_{0.8}$ films (1.1~$\mu_\mathrm{B}/$Mn in the sputtered
samples) and turned to be somewhat smaller than
2.6~$\mu_\mathrm{B}/$Mn of Mn$_5$Ge$_3$
polycrystals~\cite{Surgers2008}. At the same time, the isostructural
antiferromagnetic Mn$_5$Si$_3$ with $T_\mathrm{N}$=98~K was reported
to exhibit ferromagnetism when doped with carbon, with the transition
temperature reaching 350~K in
Mn$_5$Si$_3$C$_{0.8}$~\cite{Surgers2003}.

  While in Mn$_5$Si$_3$C$_x$ films the interstitial carbon leads to a
  lattice expansion, the increase of $T_\mathrm{C}$ upon doping in
  Mn$_5$Ge$_3$C$_x$ is accompanied by a lattice
  compression~\cite{Gajdzik2000}. However, it was suggested that the
  observed lattice distortions in both Mn$_5$Ge$_3$C$_x$ and
  Mn$_5$Si$_3$C$_x$ can have a very small influence on the increase of
  the transition temperatures~\cite{Gajdzik2000}. On the other hand,
  it was also concluded that the double-exchange mechanism plays only
  a minor role in enhancing the ferromagnetism in
  Mn$_5$Ge$_3$C$_x$. In spite of the existing experimental data on the
  magnetic properties of C-doped Mn-Ge and Mn-Si compounds, the
  information about the local magnetic moments, magnetic interactions
  and the role of carbon interstitials on the enhancement of the
  $T_\mathrm{C}$ of Mn$_5$Ge$_3$C$_x$ is still lacking. Hence, in this
  letter we explore possible zero- and finite-temperature magnetic
  properties and transition temperatures of the pure and C-doped
  Mn$_5$Ge$_3$ compounds by using first-principles calculations and
  Monte Carlo simulations.

\label{sec:method}

We follow a standard scheme for the calculation of thermodynamical
properties of magnetic systems: (1) first the Green's function
formalism and the magnetic force theorem \cite{Liechtenstein1987} are
employed to determine the exchange integrals $J_{ij}$ from first
principles; (2) these are used as input to a classical Heisenberg
Hamiltonian of the form $H=-\sum_{i,j;i\neq
  j}{J_{ij}\vec{e}_i\vec{e}_j}$, where $i$ and $j$ are the site
indexes and $\vec{e}_i$ is a unit vector along a spin moment at
$i$. The $T_\mathrm{C}$ is estimated from the peak in
susceptibility-temperature curve, obtained by Monte Carlo simulations
using 2160 magnetic-atom supercells.  The \textit{ab initio}
calculations are performed by the full-potential screened
Korringa-Kohn-Rostoker (KKR) Green function method~\cite{Ebert} within
the local spin density approximation\cite{Vosko1980} of density
functional theory. For structural relaxation we use the
projector-augmented wave approach as implemented in the ABINIT
code,\cite{ABINIT} within the generalized gradient approximation for
the exchange-correlation energy.\cite{Perdew1996}

\label{sec:analysis_discussion}

\begin{figure}
 \centering
\includegraphics[width=0.49\hsize]{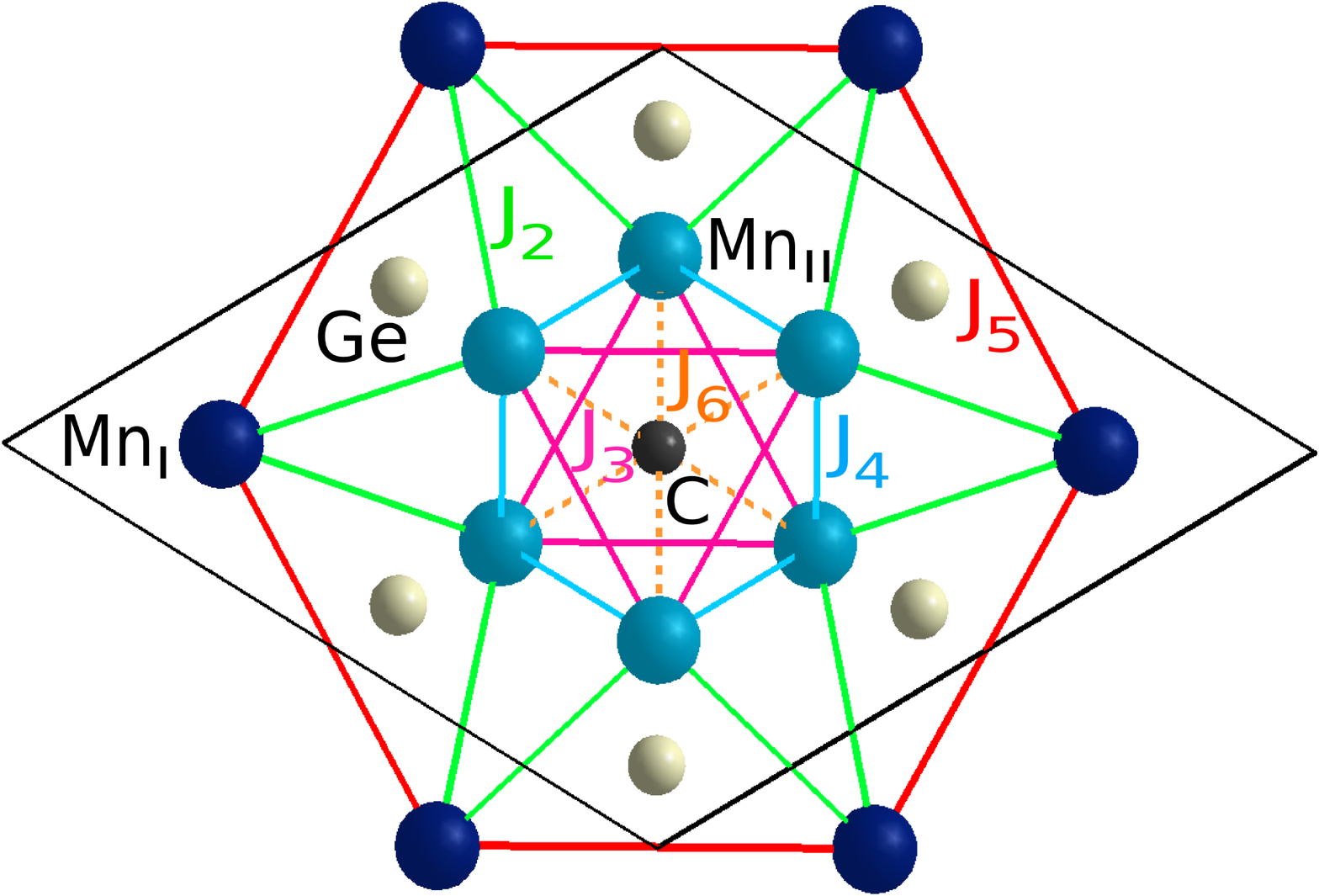}
\includegraphics[width=0.49\hsize]{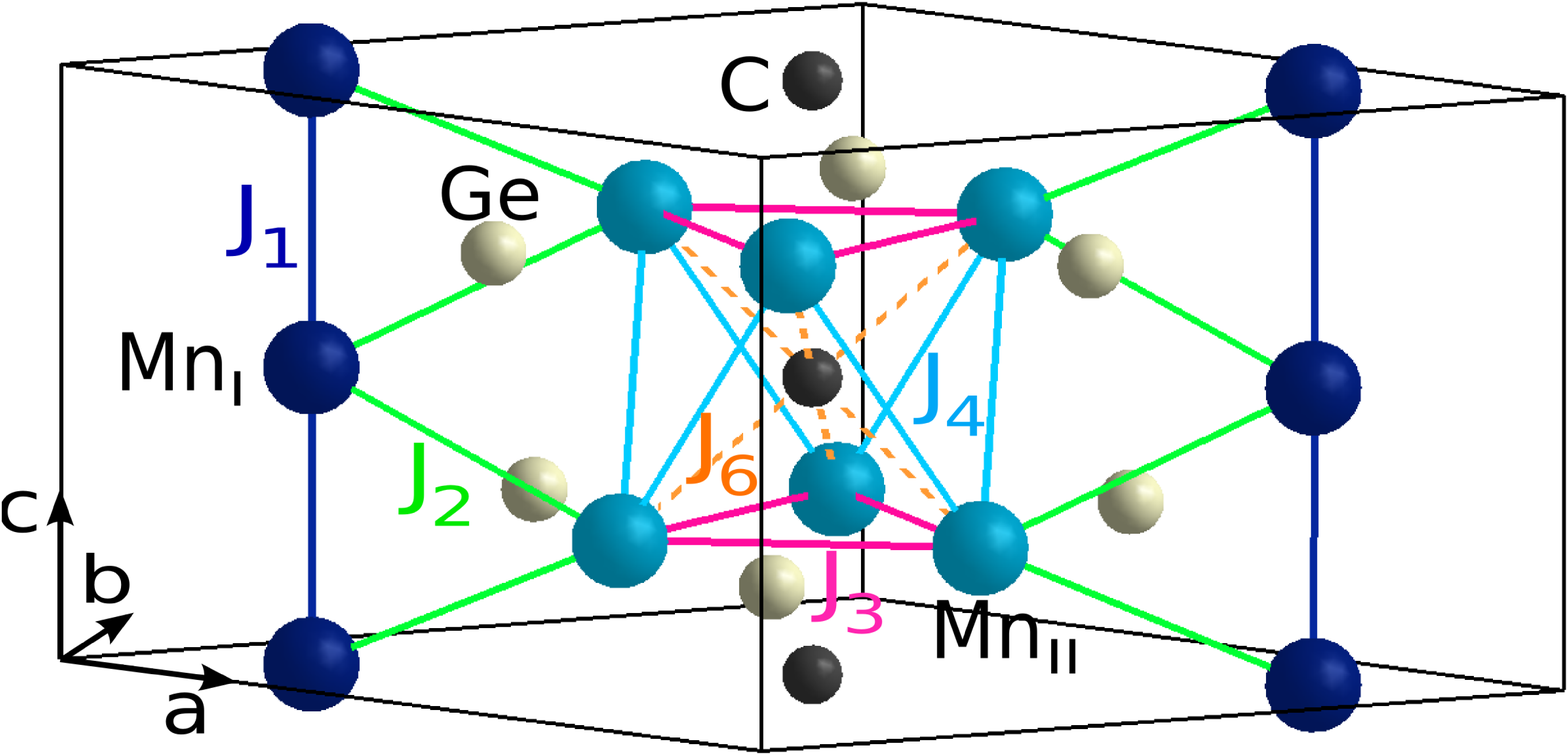}
\caption{\label{fig:Fig_1}
(Color online) Crystal structure and exchange coupling scheme for the carbon-interstitial phase Mn$_5$Ge$_3$C$_x$ at $x=1$ (full occupancy of carbon): projection on $xy$-plane (\textit{left}),
side view (\textit{right}). Colored spheres denote Ge (grey), Mn$_{\mathrm{I}}$ (dark blue), Mn$_{\mathrm{II}}$ (light blue) and interstitial C (black) at the center of Mn$_{\mathrm{II}}$ octahedra. Important exchange parameters for different Mn-Mn distances are shown in different colors. The black line indicates the edges of the unit cell.}
\end{figure}

In order to study the effect of structure relaxations on the magnetic
properties of Mn$_5$Ge$_3$ and Mn$_5$Ge$_3$C$_x$, we performed
calculations for both \textit{rigid} and \textit{relaxed} lattices.
We have used the following lattice parameters for Mn$_5$Ge$_3$, as
well as non-relaxed C-doped Mn$_5$Ge$_3$: $a=7.184$~\AA, $c/a=0.703$
\cite{Castelliz1953}. The structure of the Mn$_5$Ge$_3$C$_x$ system at
$x=1.0$ was relaxed at the experimentally obtained parameters of
Mn$_5$Ge$_3$C$_{0.75}$ films~\cite{Gajdzik2000}: $a=7.135$~\AA,
$c/a=0.700$.
The unit cell of the hexagonal Mn$_5$Ge$_3$ (space group $P6_3/mcm$,
D$8_8$-type structure ~\cite{Castelliz1953}) contains two sublattices
of Mn (Mn$_{\mathrm{I}}$ and Mn$_{\mathrm{II}}$) with different
coordination. The crystalline structure of the Mn$_5$Ge$_3$C$_x$
(Fig.~\ref{fig:Fig_1}) was found to be similar to the Mn$_5$Ge$_3$
structure with carbon atoms occupying interstitial positions at the
center of Mn$_{\mathrm{II}}$ octahedron~\cite{Gajdzik2000}.

\begin{figure}
 \centering
\includegraphics[width=0.45\hsize]{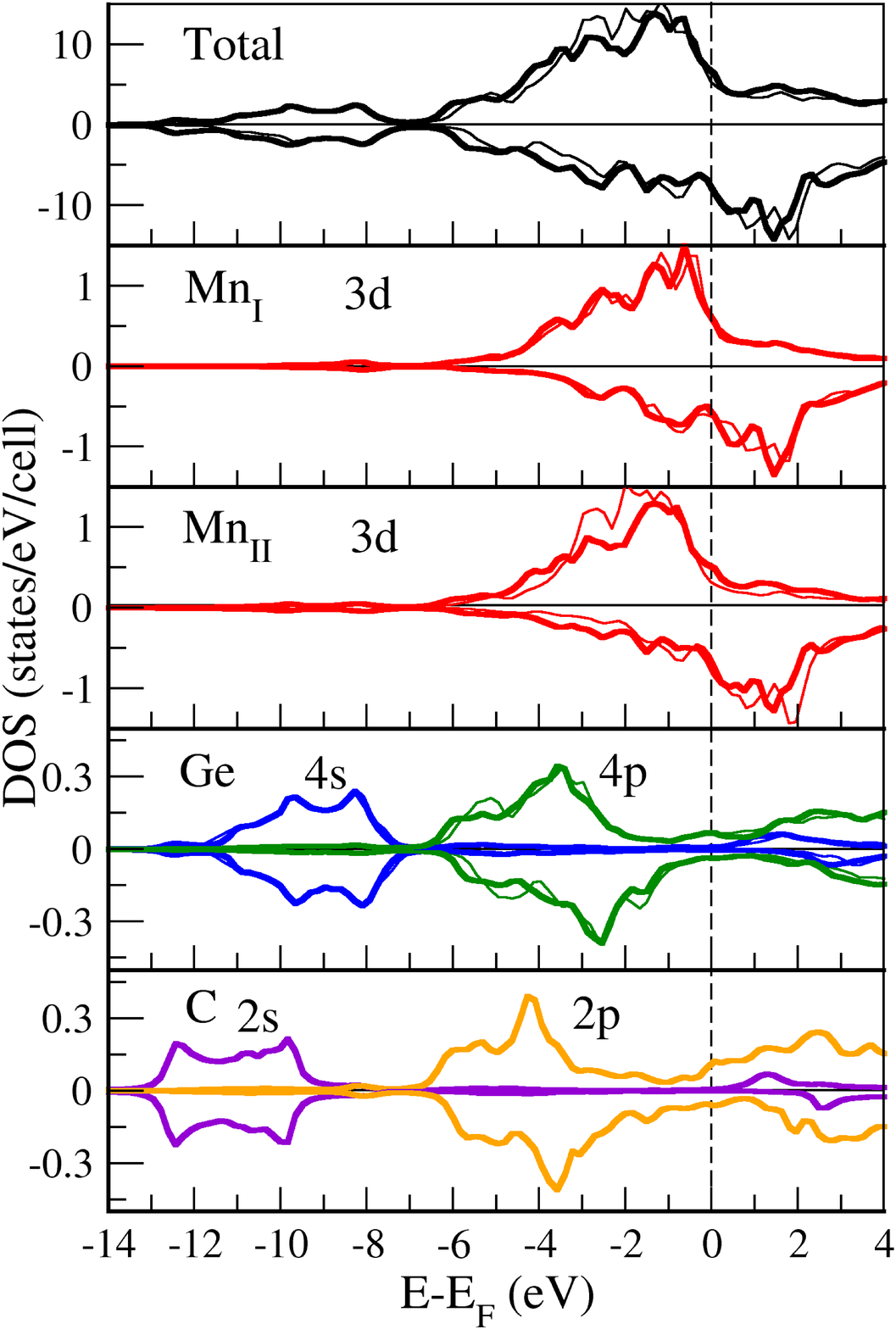}
\caption{\label{fig:Fig_2}
(Color online) Spin-resolved total and atom-projected DOS, calculated within the KKR method, for the Mn$_5$Ge$_3$ (\textit{thin line}) and Mn$_5$Ge$_3$C (\textit{thick line}) phases in FM configuration. Note the additional fine structure in the DOS of Mn$_\mathrm{II}$ for Mn$_5$Ge$_3$C due to Mn$_\mathrm{II}$-C hybridization.}
\end{figure}

According to neutron scattering experiments~\cite{Forsyth1990}, the
difference in local environment of Mn$_{\mathrm{I}}$ and
Mn$_{\mathrm{II}}$ atoms in Mn$_5$Ge$_3$ is believed to be responsible
for the different magnetic moments on them. The calculated spin
moments on Mn$_{\mathrm{I}}$, Mn$_{\mathrm{II}}$ and Ge atoms in
\textit{relaxed} (\textit{rigid}) Mn$_5$Ge$_3$ are equal to
$2.11(2.09)$, $3.11(3.15)$ and $-0.14(-0.15)$~$\mu_\mathrm{B}$
correspondingly and are almost uneffected by structural relaxation.
The smaller magnetic moment on Mn$_{\mathrm{I}}$ is due to a direct
Mn$_{\mathrm{I}}$-Mn$_{\mathrm{I}}$ interaction at a rather short
distance (2.526~\AA).\cite{Forsyth1990} The obtained values agree well
with the previous calculations\cite{Picozzi2004,Stroppa2006} and
experimental observations\cite{Forsyth1990}. The electronic density of
states (DOS) (Fig.~\ref{fig:Fig_2}) is metallic, in agreement with the
recent calculations.\cite{Picozzi2004} The partial DOS of Mn atoms is
dominated by the $3d$-states, while the largest contribution to Ge DOS
is due to $4p$- and $4s$- states. The DOS at the Fermi level
($E_{\mathrm{F}}$) is dominated by the Mn $d$-states, pointing at the
significant Mn-Mn interaction.\cite{Picozzi2004} The exchange
splitting is larger for Mn$_{\mathrm{II}}$ $d$-states compared to
Mn$_{\mathrm{I}}$, in agreement with the larger magnetic
moment on Mn$_{\mathrm{II}}$.

In order to understand the role of C doping in the enhancement of
$T_{\mathrm{C}}$, we first calculated the structurally
\textit{non-relaxed} Mn$_5$Ge$_3$C unit cell with two carbon atoms in
the octahedral voids and compared it to the corresponding results for
the undoped compound. Such situation represents the pure hybridization
effect between the carbon and the neighboring Mn$_{\mathrm{II}}$
atoms. As it follows from Fig.~\ref{fig:Fig_2}, carbon incorporation
into Mn$_5$Ge$_3$ essentially changes the 3$d$-states of
Mn$_{\mathrm{II}}$, leaving the corresponding Mn$_{\mathrm{I}}$ states
almost unaffected. The hybridization between the C $2p$- and
Mn$_\mathrm{II}$ $3d$-states leads to a shift of the main occupied
peaks in the majority, as well as unoccupied peaks in the minority
$3d$ DOS of Mn$_{\mathrm{II}}$ towards $E_{\mathrm{F}}$, consequently
increasing the density of states $N(E_{\mathrm{F}})$ at the
$E_{\mathrm{F}}$ in both spin channels. The exchange splitting is now
lower than in the pure Mn$_5$Ge$_3$, resulting in a reduced magnetic
moment on Mn$_\mathrm{II}$. The magnetic moment of the cell is mainly
contributed by the Mn$_{\mathrm{I}}$ and Mn$_{\mathrm{II}}$ atoms
($\mu_{\mathrm{Mn_{I}}}=2.21~\mu_\mathrm{B}$,
$\mu_{\mathrm{Mn_{II}}}=2.37~\mu_\mathrm{B}$), while the induced
magnetic moments on Ge and C
($\mu_{\mathrm{Ge}}=-0.14~\mu_\mathrm{B}$,
$\mu_{\mathrm{C}}=-0.26~\mu_\mathrm{B}$) are much smaller and are
antiparallel to the moments on Mn.  The structure relaxations caused
by interstitial atoms decrease the interatomic Mn-Mn distances
(Tab.~\ref{tab:table_1}), reducing the magnetic moments to
$\mu_{\mathrm{Mn_{I}}}=1.99~\mu_\mathrm{B}$,
$\mu_{\mathrm{Mn_{II}}}=2.14~\mu_\mathrm{B}$,
$\mu_{\mathrm{Ge}}=-0.13~\mu_\mathrm{B}$ and
$\mu_{\mathrm{C}}=-0.21~\mu_\mathrm{B}$. However, the effect of
hybridization on the decrease of the total magnetisation is much
larger than the effect of relaxation. This conclusion follows from our
electronic structure calculations for the relaxed Mn$_5$Ge$_3$C system
with the empty spheres at carbon positions (further referred to as
Mn$_5$Ge$_3$V$_\mathrm{C}$). The calculated magnetic moments on Mn
agree well with the average saturated moment of
2.2~$\mu_\mathrm{B}/$Mn, observed for the C-implanted
Mn$_5$Ge$_3$C$_{0.8}$ films,\cite{Surgers2008} but deviate from
those obtained for the sputtered samples, where the largest moment of
1.1~$\mu_\mathrm{B}/$Mn was observed for Mn$_5$Ge$_3$C$_{0.75}$
\cite{Gajdzik2000}. This discrepancy could be ascribed to disorder
effects and defects formation, which may differ in sputtered and
implanted samples.

In Tab.~\ref{tab:table_1} we present some of the important calculated
exchange constants for the relaxed and rigid Mn$_5$Ge$_3$ and
Mn$_5$Ge$_3$C systems (see Fig.~\ref{fig:Fig_1} on the notation). Only the
interactions between Mn spins are shown, as Mn-Ge interactions turned
to be negligibly small. The first nearest neighbor
Mn$_{\mathrm{I}}$-Mn$_{\mathrm{I}}$ interactions $J_1$ (at the
shortest distance of 2.526~\AA) are ferromagnetic (FM) and clearly dominate
over the corresponding Mn$_{\mathrm{I}}$-Mn$_{\mathrm{II}}$ and
Mn$_{\mathrm{II}}$-Mn$_{\mathrm{II}}$ interactions, confirming the
assumption about the different magnitude of exchange interaction for
Mn atoms in different sublattices \cite{Panissod1984}. The dependence
of the exchange parameters on the interatomic distances is stronger
for the Mn$_{\mathrm{II}}$ sublattice, while it is weaker for the
Mn$_{\mathrm{I}}$ one.
As it follows from Tab.~\ref{tab:table_1}, there is an
antiferromagnetic (AFM) $J_3$ interaction between the
Mn$_{\mathrm{II}}$ atoms within $z=\frac{1}{4}$ and $z=\frac{3}{4}$
planes (perpendicular to the $c$-axis), and FM $J_4$
and $J_6$ interactions between Mn$_{\mathrm{II}}$ belonging to
different planes. The negative sign for $J_3$ could be responsible for
the small degree of non-collinearity in Mn$_5$Ge$_3$, observed in the
previous first-principles calculations \cite{Stroppa2006}.
However, an increase of the corresponding Mn$_{\mathrm{II}}$-Mn$_{\mathrm{II}}$
distance from 2.974~\AA\ to 3.057~\AA\ after structure relaxation
leads to the change from AFM to FM $J_3$ parameter, while the rest of
the parameters are left almost unchanged. As a result the estimated
value of the $T_{\mathrm{C}}=400$~K for the \textit{relaxed}
Mn$_5$Ge$_3$ system is somewhat higher than for the \textit{rigid} one
of 320~K, the latter being in a very good agreement with the
corresponding experimentally observed value of 304~K
\cite{Tawara1963}.

The results presented in Tab.~\ref{tab:table_1} show that interstitial
carbon, even without relaxation, significantly changes the value of
$J_2$, $J_3$ and $J_6$, leaving the $J_4$ and $J_1$
interactions almost uneffected. In particular, the $J_3$ interaction
becomes strongly FM compared to the undoped Mn$_5$Ge$_3$. According to
the DOS of Mn$_5$Ge$_3$C in Fig.~\ref{fig:Fig_2}, this increase of the
FM interaction can probably be explained by the strong $p$-$d$
hybridization between Mn$_{\mathrm{II}}$ and C states, which enhances
the hopping of the 3$d$ electrons from the partially occupied
$d$-states of one Mn$_{\mathrm{II}}$ atom to a 3$d$ orbital of the
neighboring Mn$_{\mathrm{II}}$. The $J_6$ parameter, which in the
doped compound corresponds to the 180$^{\circ}$
Mn$_{\mathrm{II}}$-C-Mn$_{\mathrm{II}}$ interaction (\textit{dashed}
line on Fig.~\ref{fig:Fig_1}), is essentially lower than the
corresponding direct Mn$_{\mathrm{II}}$-Mn$_{\mathrm{II}}$ exchange in
the parent compound, pointing to the possible superexchange
interaction between Mn$_{\mathrm{II}}$ via carbon, as predicted by
Goodenough-Anderson-Kanamori (GAK) rules.\cite{Goodenough1963}
According to these, 90$^{\circ}$
Mn$_{\mathrm{II}}$-C-Mn$_{\mathrm{II}}$ interactions should be FM,
which is the case for $J_3$ and $J_4$. In this model interatomic
exchange constants are strongly dependent on the orbital overlap, and
hence sensitive to changes in the interatomic distances. It should be
noted that GAK rules have been succesfully applied to explain the
first-order FM-to-AFM transition in Mn$_3$GaC$_{1-\delta}$
antiperovskite compounds,\cite{Lewis2006} in which the local
environment of Mn atoms is similar to that in Mn$_5$Ge$_3$C.

The transition temperature, calculated using the exchange parameters
for the \textit{non-relaxed} C-doped system, is essentially
higher ($T_{\mathrm{C}}=430$~K) than in the undoped compound. This is
a direct indication that the strongly enhanced FM stability is not
related solely to the variation of the Mn-Mn interatomic distances. To
distinguish the effect of relaxations from the electronic structure
effect on the enhanced FM interaction in Mn$_5$Ge$_3$C, we compared
the values of the exchange parameters in the
Mn$_5$Ge$_3$V$_\mathrm{C}$ with those calculated for the parent
compound. We can conclude that the essential decrease of the
Mn$_{\mathrm{II}}$-Mn$_{\mathrm{II}}$ distances due to carbon
insertion reduces the FM exchange interaction, and has a particular
influence on  $J_3$  making it strongly AFM, thus
destabilizing the collinear magnetic structure in
Mn$_5$Ge$_3$V$_\mathrm{C}$ (Tab.~\ref{tab:table_1}). The relaxation
makes  $J_3$  in Mn$_5$Ge$_3$C somewhat lower, but still
high compared to that in the pure compound. The $J_4$ parameter
is higher in \textit{relaxed} Mn$_5$Ge$_3$C, pointing to the
90$^{\circ}$ FM superexchange, enhanced by the corresponding distance
decrease. Although the strong hybridization between Mn $3d$ and C $2p$
orbitals leads to the contraction of the interatomic distances and
lowers the longer-distance exchange constants, the FM Mn-Mn
interactions prevail in the Mn$_5$Ge$_3$C, thus giving larger
contribution to the $T_\mathrm{C}$ than the AFM ones. The
calculated $T_\mathrm{C}$=450~K is only 20~K larger than in the
non-relaxed Mn$_5$Ge$_3$C, pointing to the small effect of the
structural distortions on the $T_\mathrm{C}$ increase  in
Mn$_5$Ge$_3$ due to C-doping. The obtained value is in a very
good agreement with the maximum value of the experimentally measured
$T_\mathrm{C}$ of 445~K\cite{Gajdzik2000,Surgers2008}. To summarize, the present 
studies clarified that the enhanced FM stability of C-doped Mn$_5$Ge$_3$ 
compounds is attributed to the appearance of the 90$^{\circ}$ FM superexchange 
between Mn atoms mediated by carbon, while the structural distortions are found 
to be a secondary effect. \\

\label{sec:acknowledgements}

I.S. acknowledges the CEA nanosciences program and P.M. the ESF SONS, Contract ERAS-CT-2003-980409, for partial funding
The simulations were performed at the CEA supercomputing center (CCRT)
and at the J\"ulich Supercomputing center in the framework of the "High Performance Computing" collaboration between CEA and Helmholtz institutes.

\label{sec:bibliography}
%

\newpage

\begin{thebibliography}{10}


\bibitem{Picozzi2004}
S.~{Picozzi}, A.~{Continenza}, and A.~J. {Freeman},
\newblock {Phys. Rev. B} \textbf{70}, 235205 (2004).

\bibitem{Tawara1963}
Y.~{Tawara} and K.~{Sato},
\newblock {J. Phys. Soc. Jpn.} \textbf{18}, 773 (1963).

\bibitem{Gajdzik2000}
M.~{Gajdzik}, C.~{S{\"u}rgers}, M.~T. {Kelemen}, and H.~v. {L{\"o}hneysen},
\newblock {J. of Magn. Magn. Mat.} \textbf{221}, 248 (2000).

\bibitem{Surgers2008}
C.~{S{\"u}rgers}, K.~{Potzger}, T.~{Strache}, W.~{M{\"o}ller}, G.~{Fischer},
  N.~{Joshi}, and H.~{v.~L{\"o}hneysen},
\newblock {Appl. Phys. Lett.} \textbf{93}, 062503 (2008).

\bibitem{Surgers2003}
C.~{S{\"u}rgers}, M.~{Gajdzik}, G.~{Fischer}, H.~V. {L{\"o}hneysen},
  E.~{Welter}, and K.~{Attenkofer},
\newblock {Phys. Rev. B} \textbf{68}, 174423 (2003).

\bibitem{Liechtenstein1987}
A.~I. {Liechtenstein}, M.~I.~{Katsnelson}, V.~P.~{Antropov}, and V.~A.~{Gubanov},
\newblock {J. Magn. Magn. Mater.} \textbf{67}, 65 (1987).


\bibitem{Ebert}
{The SPR-TB-KKR package, H. Ebert and R. Zeller,~http: //olymp.cup.uni-muenchen.de/ak/ebert/SPR-TB-KKR}.

\bibitem{Vosko1980}
S.~H. {Vosko}, L.~{Wilk}, and M.~{Nusair},
\newblock {Canad. J. Phys.} \textbf{58}, 1200 (1980).


\bibitem{ABINIT}
X.~{Gonze} et al., {Comp. Mat. Sci.} \textbf{25}, 478 (2002).

\bibitem{Perdew1996}
J.~P. {Perdew}, K.~{Burke}, and M.~{Ernzerhof},
\newblock {Phys. Rev. Lett.} \textbf{77}, 3865 (1996).

\bibitem{Castelliz1953}
L.~{Castelliz},
\newblock {Mh. Chem.} \textbf{84}, 765 (1953).

\bibitem{Forsyth1990}
J.~B. {Forsyth} and P.~J. {Brown},
\newblock {J. Phys.: Condens. Matter.} \textbf{2}, 2713 (1990).

\bibitem{Stroppa2006}
A.~{Stroppa} and M.~{Peressi},
\newblock {Phys. Stat. Sol. (a)} \textbf{204}, 44 (2006).

\bibitem{Panissod1984}
P.~{Panissod}, A.~{Qachaou}, and G.~{Kappel},
\newblock {J. Phys. C: Sol. St. Phys.} \textbf{17}, 5799 (1984).

\bibitem{Goodenough1963}
J.B. {Goodenough},
\newblock {Magnetism and the Chemical Bond, New York: Interscience, 1963}.

\bibitem{Lewis2006}
L.H. {Lewis}, D.~{Yoder}, A.R. {Moodenbaugh}, D.A. {Fischer}, and M-H. {Yu},
\newblock {J. Phys.: Condens. Matter.} \textbf{18}, 1677 (2006).

\end{thebibliography}

%

%
\newpage
\mbox{}

\newpage
\mbox{}
\newpage

\begin{table*}
 \caption{Calculated (within KKR) exchange constants $J_{ij}$  and $T_{\mathrm{C}}$ for the rigid and relaxed Mn$_5$Ge$_3$ and Mn$_5$Ge$_3$C compounds. The corresponding Mn-Mn distances (in~\AA) are shown in square brackets. Positive (negative) values characterize FM (AFM) coupling. Results for Mn$_5$Ge$_3$V$_\mathrm{C}$ are presented to distinguish between the structural and chemical effects of doping.
}\label{tab:table_1}
\begin{ruledtabular}
   \begin{tabular}{c|ccccc}
    $J_{ij}$ (mRy) & Mn$_5$Ge$_3$ (rigid) & Mn$_5$Ge$_3$ (relaxed) & Mn$_5$Ge$_3$C (rigid) & Mn$_5$Ge$_3$C (relaxed) & Mn$_5$Ge$_3$V$_\mathrm{C}$ \\
    \hline\
    $J_{1}^{\mathrm{Mn_I}-\mathrm{Mn_I}}$       &  $+2.14$ [2.526]  &   $+2.09$ [2.527]  & $+2.24$ [2.526]  & $+2.34$ [2.498] & $+2.24$ [2.498]\\
    $J_{2}^{\mathrm{Mn_I}-\mathrm{Mn_{II}}}$    &  $+0.59$ [3.068]  &   $+0.59$ [3.039]  & $+0.90$ [3.068]  & $+0.69$ [3.124] & $+0.40$ [3.124]\\
    $J_{3}^{\mathrm{Mn_{II}}-\mathrm{Mn_{II}}}$ &  $-0.15$ [2.974]  &   $+0.13$ [3.057]  & $+0.84$ [2.974]  & $+0.51$ [2.714] & $-1.80$ [2.714]\\
    $J_{4}^{\mathrm{Mn_{II}}-\mathrm{Mn_{II}}}$ &  $+0.51$ [3.055]  &   $+0.52$ [3.082]  & $+0.44$ [3.055]  & $+0.60$ [2.948] & $+0.28$ [2.948]\\
    $J_{5}^{\mathrm{Mn_I}-\mathrm{Mn_{I}}}$     &  $-0.10$ [4.148]  &   $-0.06$ [4.148]  & $-0.16$ [4.148]  & $-0.22$ [4.118] & $-0.09$ [4.118]\\
    $J_{6}^{\mathrm{Mn_{II}}-\mathrm{Mn_{II}}}$ &  $+0.69$ [4.263]  &   $+0.65$ [4.341]  & $+0.07$ [4.263]  & $+0.22$ [4.008] & $+0.76$ [4.008]\\
    \hline\
    $T_{\mathrm{C}}$ (K) &  320 &   400   &  430 & 450  & non-collinear \\
    \end{tabular}
\end{ruledtabular}
\end{table*}

\end{document}